\documentclass[a4paper,12pt,openany]{article}

\usepackage{color}
\definecolor{darkblue}{rgb}{0.1,0.1,.7}
\usepackage[colorlinks, linkcolor=darkblue, citecolor=darkblue, urlcolor=darkblue, linktocpage]{hyperref}

\usepackage[]{amsmath}
\usepackage[]{graphicx}
\usepackage[]{latexsym}
\usepackage{slashed,graphicx,color,amsmath,amssymb}
\usepackage[mathscr]{eucal}
\usepackage{mathrsfs}
\usepackage{geometry}
\usepackage[margin=10pt,font=small,labelfont=bf]{caption}
\usepackage{amscd}
\usepackage{bm}
\usepackage{xcolor}
\usepackage{lipsum}
\usepackage{upgreek}
\numberwithin{equation}{section}
\newcommand{\tr}{\mathrm{Tr}\,}

\newcommand{\be}{\begin{equation}}
	\newcommand{\ee}{\end{equation}}
\newcommand{\bea}{\begin{eqnarray}}
	\newcommand{\eea}{\end{eqnarray}}
\newcommand{\ii}{\mathrm{i}}
\title{\Large Three-point Functions in Aharony-Bergman-Jafferis-Maldacena Theory and Integrable Boundary States}
\author{\small
	Jun-Bao Wu\textsuperscript{a, b}\thanks{\href{mailto:junbao.wu@tju.edu.cn}{junbao.wu@tju.edu.cn}} \and \small
	Peihe Yang\textsuperscript{c, b}\thanks{\href{mailto:peiheyang@pku.edu.cn}{peiheyang@pku.edu.cn}}
}

\date{\small
	\textsuperscript{a}Center for Joint Quantum Studies and Department of Physics, School of Science, Tianjin University, 135 Yaguan Road, Tianjin 300350, P. R. China \\
	\textsuperscript{b}Peng Huanwu Center for Fundamental Theory, 1 Xuefu Avenue, Xi'an, 710127, P. R. China \\
	\textsuperscript{c}Center for High Energy Physics, Peking University, 5 Yiheyuan Road, Beijing 100871, China \\
}

\begin{document}
\maketitle
\begin{abstract}
		We investigate the correlators of three single-trace operators in Aharony–Bergman-\\Jafferis–Maldacena (ABJM) theory from the perspective of integrable boundary states. Specifically, we focus on scenarios where two operators are $1/3$-BPS and the entire correlation function is considered within the twisted-translated frame. The correlator can be expressed as the overlap between a boundary state and a Bethe state. It is found that the boundary state formed by the two $1/3$-BPS operators is integrable when the number of Wick contractions between the non-BPS operator and one of the $1/3$-BPS operators is $0$ or $1$. We compute the overlaps for the proven integrable cases utilizing the symmetries and the coordinate Bethe ansatz.
	\end{abstract}
	\newpage

\tableofcontents
    
	\section{Introduction}
 In conformal field theories, two-point and three-point functions (structure constants) of local operators determine higher-point functions and are collectively known as the CFT data~\cite{Simmons-Duffin:2016gjk}. In general gauge theories, computing conformal dimensions and structure constants of single trace operators at arbitrary coupling is challenging. Integrable structures~\cite{Beisert:2010jr} in planar ${\mathcal N}=4$ super Yang-Mills (SYM) and Aharony–Bergman–Jafferis–Maldacena (ABJM) theories~\cite{Aharony:2008ug} have made substantial contributions to solving these problems. The quantum spectral curve method solves the spectral problem at any 't Hooft coupling~\cite{Gromov:2013pga,Gromov:2014caa,Cavaglia:2014exa,Bombardelli:2017vhk}, and the hexagon program provides a framework for computing the three-point functions in ${\mathcal N}=4$ SYM~\cite{Basso:2015zoa}~\footnote{For recent progress involving operators of any length, see~\cite{Basso:2022nny}.}. In ABJM theory, while some hexagonal form factors were computed~\cite{Pereira:2017unx}, three-point functions of single-trace operators are much less explored in the framework of AdS/CFT integrability. To the best of our knowledge, \textit{only} three-point functions in the $SU(2) \times SU(2)$ sector
	were studied using integrability~\cite{Bissi:2012ff}. In this paper, we aim to study a special three-point functions in the full scalar sector from the perspective of \textit{integrable boundary states}.
	
	In AdS/CFT integrability, integrable boundary states~\cite{Piroli:2017sei} appear in one-point functions with non-trivial backgrounds (such as the presence of a domain wall~\cite{deLeeuw:2015hxa, Kristjansen:2021abc}, a BPS Wilson loop~\cite{JKV, Jiang:2023cdm} or a 't~Hooft loop~\cite{Kristjansen:2023ysz}, or in the Coulomb branch~\cite{Ivanovskiy:2024vel}) and in three-point functions involving two BPS determinant operators~\cite{Jiang:2019xdz, Jiang:2019zig, Yang:2021hrl}. These correlation functions are expressed as overlaps between an integrable boundary state and a Bethe state. It is to be explored whether the correlation functions containing three single-trace operators will provide new integrable boundary states. This paper investigates tree-level three-point functions with two $1/3$-BPS operators in ABJM theory, and searches the condition for the boundary state from these two BPS operators to be integrable.  We  find that the above boundary state is integrable  when the correlator is special extermal or special next-to-extremal~\footnote{By special (next-to-)extremal correlators, we mean that all (but one) Wick contractions involving the non-BPS operator are with one of the two $1/3$-BPS operators. }. It turns out that the integrable states in the latter case are of a novel  type.  We further require that this three-point function is in the twisted-translated frame~\cite{Drukker:2009sf, Basso:2015zoa, Pereira:2017unx, Yang:2021hrl}. By utilizing the constraints from conformal and  R symmetries~\cite{Kazama:2014sxa, Yang:2021hrl}, along with the coordinate Bethe ansatz, we compute the correlation function analytically for the known integrable cases. 

The outline of the remaining parts of the paper is as follows.
%In section~\ref{Section:notations}, we briefly review the basic notations of the $SU(4)$ alternating spin chain and  the integrable boundary states. 
In section~\ref{Section:single-trace}, we  review the single-trace operators in ABJM theory and some properties  of their correlation functions. In the next two sections, we construct the boundary states from the Wick contractions for certain three-point functions and find that these states are integrable for special (next-to-)extremal correlators. Finally in section~\ref{Section:overlaps}, we compute the correlators for proven integrable cases.  In section~\ref{outlook}, we summarize our  results and discuss possible further directions. In Appendix~\ref{appendix3}, we provide criteria for integrable boundary states in the $SU(N)$ spin chain using an approach similar to the one in section~\ref{Section:integrability}. We put some basic notations and technical details in the other three appendices.

  \section{Single-trace operators and their correlation functions}\label{Section:single-trace}

 ABJM theory is a three-dimensional ${\mathcal {N}}=6$ super-Chern-Simons theory with gauge group $U(N)\times U(N)$ and Chern-Simons level $k$ and $-k$, respectively. Among the matter fields, there are four complex scalars $Y^I, I=1, \cdots, 4$ in the bifundamental representation of the gauge group and the fundamental representation of the $SU(4)$ R-symmetry group.

	The  length-$2L$ single-trace operator in the scalar sector of the ABJM theory is %\footnote{In the following, the single-trace operators are denoted by $\hat{\mathcal{O}}$, except for those appearing in the proof of the integrable boundary state.} 
	\be \hat{\mathcal{O}}_C=C_{I_1\cdots I_L}^{J_1\cdots J_L}{\mathrm{Tr}}(Y^{I_1}Y^\dagger_{J_1}\cdots Y^{I_L}Y^\dagger_{J_L})\,. \ee
	The cyclicity property of the trace can be used to choose $C$ to be invariant under the following simultaneous cyclic shift  of the upper and lower indices, $C_{I_1\cdots I_L}^{J_1 \cdots J_L}=C_{I_2\cdots I_L I_1}^{J_2 \cdots J_L J_1}$.

Such single BPS operator becomes $1/3$-BPS when the tensor $C$ is invariant under the respective permutations among the upper or the lower indices, and  traceless, 
	\be  C_{I_1\cdots I_L}^{J_1\cdots J_L}=C_{(I_1\cdots I_L)}^{J_1\cdots J_L}=C_{I_1\cdots I_L}^{(J_1\cdots J_L)}\,,\, C_{I_1\cdots I_L}^{J_1\cdots J_L}\delta_{J_1}^{I_1}=0\,,\ee
		A natural choice of such symmetric traceless tensor $C$ is in terms of polarization vectors $n_I$ and $\bar{n}^I$~\footnote{Notice that $\bar{n}$ does not need to be the complex conjugation of $n$.},
	\be C_{I_1\cdots I_L}^{J_1\cdots J_L}=n_{I_1}\cdots n_{I_L}\bar{n}^{J_1}\cdots \bar{n}^{J_L}\,,\ee
	with BPS condition $n\cdot \bar{n}=0$.
	With this choice, the BPS operator becomes
	\be \hat{\mathcal{O}}^\circ_{L}(x, n, \bar{n})=\tr \left(\left(n\cdot Y \bar{n}\cdot Y^\dagger\right)^L\right).\, \ee
 Here the superscript $^\circ$ is used to indicate that the operator is BPS. 
	
	The two-point function of $\hat{\mathcal{O}}^\circ_{L}$'s is constrained by  symmetries to take the form,
	\be \langle \hat{\mathcal{O}}_{L_1}^\circ(x_1)\hat{\mathcal{O}}_{L_2}^\circ(x_2)\rangle=\delta_{L_1, L_2} \mathcal{N}_{\hat{\mathcal{O}}^\circ_{L_1}}(d_{12}d_{21})^{L_1}, \ee
	with the definitions 
	\be d_{ij}=\frac{n_i\cdot \bar{n}_j}{|x_{ij}|}\,, \,
	x_{ij}=x_i-x_j\,.\ee
	At planar tree-level, we have 
	\be\mathcal{N}_{\hat{\mathcal{O}}^\circ_{L}}=L\lambda^{2L}\,.\label{normBPS} \ee
	where $\lambda:= N/k $ is the \textquoteright{}t~Hooft coupling constant of ABJM theory.
	
	For a non-BPS operator $\hat{\mathcal{O}}$ in the scalar sector, the normalization $\mathcal{N}_{\hat{\mathcal{O}}}$ is defined by means of the two-point function of this operator and its Hermitian conjugate $\hat{\mathcal{O}}^\dagger$,
	\be \langle \hat{\mathcal{O}}(x) \hat{\mathcal{O}}^\dagger(0) \rangle=\frac{\mathcal{N}_{\hat{\mathcal{O}}}}{|x|^{2\Delta_{\hat{\mathcal{O}}}}} \,,\ee
	where $\Delta_{\hat{\mathcal{O}}}$ is the conformal dimension of $\hat{\mathcal{O}}$. At tree-level, we have $\Delta_{\hat{\mathcal{O}}}=L$ when the length of $\hat{\mathcal{O}}$ is $2L$.
	
	We now consider the twisted-translated frame, in which aligning all operators along a straight line constitutes part of its definition. Here we consider this line to be  the $x_3$ axis.~\footnote{We put the theory 		in the three-dimensional  Euclidean  space $\mathbf{R}^3$ with coordinates $x_\mu=(x_1, x_2, x_3)$.}.
	For the BPS operator at $x^\mu=(0, 0, a)$, we choose the polarization vectors to be 
	\be n=(1, 0, 0, \kappa a)\,,\, \bar{n}=(-\kappa a, 0, 0, 1)\,. \ee
	Here  the parameter $\kappa$ has been introduced to make $\kappa a$ dimensionless.
	Now for two BPS operators $\hat{\mathcal{O}}^\circ_1$ and $\hat{\mathcal{O}}^\circ_2$ in the twisted-translated frame, we have $d_{12}=\mathrm{sgn}(a_{12})$.
	
	The twisted-translated frame for general operators in the scalar sector is defined by performing the transformation 
	\be Y^1\to Y^1+\kappa a Y^4\,, Y^\dagger_4\to Y^\dagger_4-\kappa a Y^\dagger_1\,,\ee 
	while leaving other fields unchanged, when shifting the position of the operators from the origin to $x^\mu=(0, 0, a)$. 
	
	According to the result in \cite{Kazama:2014sxa, Yang:2021hrl}, the normalized correlation function of three generic single-trace operators in the twisted-translated frame with $\kappa=1$~\footnote{The $\kappa$-dependence can be easily recovered from dimensional analysis. So for now on, we will keep $\kappa=1$.} should take the following form,
	\be \frac{\langle \hat{\mathcal{O}}_1(a_1) \hat{\mathcal{O}}_2(a_2) \hat{\mathcal{O}}_3(a_3)\rangle}{\sqrt{\mathcal{N}_{\hat{\mathcal{O}}_1}\mathcal{N}_{\hat{\mathcal{O}}_2}\mathcal{N}_{\hat{\mathcal{O}}_3}}}=
	\frac{\mathcal{C}_{123}}{a_{12}^{\gamma_{12|3}} a_{23}^{\gamma_{23|1}}a_{31}^{\gamma_{31|2}}}\,,\label{symmetry}\ee
	where 
	\be\gamma_{pq|r}:=(\Delta_p+\Delta_q-\Delta_r)-(J_p+J_q-J_r)\,.\ee
	Here we consider  a $U(1)$ R-charge $J$ which assigns  charges $(1/2, 0, 0, -1/2)$ to $Y^1, \cdots, Y^4$.  $J_p$ is the corresponding charge of the operator $\hat{\mathcal{O}}_p, p=1, 2, 3$ at the origin (before performing the twisted-translation).~\footnote{ In other words, when we compute the charge of $\hat{\mathcal{O}}_p$ directly at $a_i$, we should assign the parameter $\kappa$ $J$-charge $1$ after restoring this parameter. } For the BPS operator $\hat{\mathcal{O}}_i, i=1, 2$, we have $J_i=L_i=\Delta_i$.  Notice that \eqref{symmetry} is valid for  three-point functions with generic $\hat{\mathcal{O}}$'s.
	
	The main focus of this paper is on  three-point functions of two $1/3$-BPS single-trace operators ${\hat{\mathcal{O}}}^\circ_{L_i}, i=1, 2$ and one non-BPS operator $\hat{\mathcal{O}}_3$.  For this special case, \eqref{symmetry} leads to 
	\be\label{constraint} \frac{\langle \hat{\mathcal{O}}_1^\circ(a_1) \hat{\mathcal{O}}_2^\circ(a_2) \hat{\mathcal{O}}_3(a_3)\rangle}{\sqrt{\mathcal{N}_{\hat{\mathcal{O}}^\circ_1}\mathcal{N}_{\hat{\mathcal{O}}^\circ_2}\mathcal{N}_{\hat{\mathcal{O}}_3}}}=\mathcal{C}_{123}\left(\frac{a_{12}}{a_{23}a_{31}}\right)^{\Delta_3-J_3} \,.\ee

 \section{Boundary states from Wick contractions}\label{Section:boundarystates}
	 In the large $N$ limit, the tree-level three-point function~\footnote{Without loss of generality, we have already set $a_3=0$. } 
	\be \langle \hat{\mathcal{O}}_1^\circ(a_1) \hat{\mathcal{O}}_2^\circ(a_2) \hat{\mathcal{O}}_3(0)\rangle\,, \ee
	is computed from planar Wick contractions, where $\sum_{i=1}^3L_i$ pairs of fields are contracted.  Let us denote the number of the contractions between operators $\hat{\mathcal{O}}_p$ and $\hat{\mathcal{O}}_q$ by $l_{pq|r}, r\neq p, q, 1\leq p, q, r\leq 3$. It is straightforward to see that
	$l_{pq|r}=L_p+L_q-L_r$.
	Notice that $l_{12|3}, l_{23|1}, l_{31|2}$ are either all even or all odd. This behavior contrasts with the one in $\mathcal{N}=4$ SYM theory.
	
	When $l_{pq|r}$'s are even and $2\leq l_{31|2} \leq 2L_3-2$,  planar Wick contractions among these three operators yield 
\begin{equation}
\begin{aligned}
		\label{WickEven} &\langle \hat{\mathcal{O}}_1^\circ(a_1) \hat{\mathcal{O}}_2^\circ(a_2) \hat{\mathcal{O}}_3(0)\rangle  
	=\frac{(-1)^{\frac{l_{12|3}}{2}}L_1L_2\lambda^{\sum_{i=1}^3L_i}}{N|a_1|^{l_{31|2}} |a_2|^{l_{23|1}}} C_{I_1\cdots I_{L_3}}^{J_1\cdots J_{L_3}}\nonumber\\
	&\times
	\sum_{1\leq s, s^\prime \leq L, \,s^\prime-s=\frac{l_{31|2}}{2}-1}
	\left((\bar{n}_2)^{I_1}(n_2)_{J_1}\cdots (n_2)_{J_{s-1}}(\bar{n}_1)^{I_s}\cdots (n_1)_{J_{s^\prime}}(\bar{n}_2)^{I_{s^\prime+1}}\cdots (n_2)_{J_{L_3}} \right.\nonumber\\&
	+ \left. (\bar{n}_2)^{I_1}(n_2)_{J_1}\cdots (\bar{n}_2)^{I_s} (n_1)_{J_s}\cdots (\bar{n}_1)^{I_{s^\prime+1}}(n_2)_{J_{s^\prime+1}}\cdots (n_2)_{J_{L_3}}\right)
	\,. 
\end{aligned}
\end{equation}
	Here and in the following, $s^\prime-s$ should be understood in the sense of $\textrm{mod} \, L_3$ and we have used that the Wick contractions between $\hat{\mathcal{O}}^\circ_1$ and  $\hat{\mathcal{O}}^\circ_2$ yield the factor $(d_{12} d_{21})^\frac{l_{12|3}}{2}=(-1)^\frac{l_{12|3}}{2}$.
	
	The cases with $l_{31|2}=0 $ or $2L_3$ are special.  In these scenarios, we have either $L_2=L_3+L_1$ or $L_1=L_2+L_3$, which makes the three-point function a special extremal one. For these cases, the computation of correlators must account for the mixing of  $\hat{\mathcal{O}}_2$ or 
	$\hat{\mathcal{O}}_1$ with double trace operators, even in the large $N$ limit~\cite{DHoker:1999jke}. Using the case with $l_{31|2}=0$ as an example, we should  find the eigen-operators
\begin{equation}
\hat{\mathcal{O}}^{\mathrm{new}}_2=\hat{\mathcal{O}}^\circ_2+
\tilde{C}^{J_1\cdots J_{L_1}J_{L_1+1}\cdots J_{L_2}}_{I_1\cdots I_{L_1}I_{L_1+1}\cdots I_{L_2}}\mathrm{Tr}(Y^{I_1}Y_{J_1}^\dagger\cdots Y^{I_{L_1}}Y_{J_{L_1}} )\mathrm{Tr}(Y^{I_{L_1+1}}Y_{J_{L_1+1}}^\dagger\cdots Y^{I_{L_2}}Y_{J_{L_2}})\,,
\end{equation}
 of two-loop dilatation operator~\footnote{Here the components of the tensor $\tilde{C}$ should be obtained by solving the eigenvalue equations.}, and compute the three-point function
 $\langle \hat{\mathcal{O}}^{\circ}_1(a_1)\hat{\mathcal{O}}^{\mathrm{new}}_2(a_2) \hat{\mathcal{O}}_3(0)\rangle$.
This correlator is certainly different from the three point functions $\langle \hat{\mathcal{O}}^{\circ}_1(a_1)\hat{\mathcal{O}}^{\circ}_2(a_2) \hat{\mathcal{O}}_3(0)\rangle$ which we originally aimed to compute. Notice that the double trace terms in $\mathcal{O}^{\mathrm{new}}_2$
prevent the correlation functions from being simply written as the overlaps between a boundary state and Bethe states on the Hilbert space of closed $SU(4)$ alternating spin chain.
    Finding $\hat{\mathcal{O}}^{\mathrm{new}}_2(a_2)$ and computing $\langle \hat{\mathcal{O}}^{\circ}_1(a_1)\hat{\mathcal{O}}^{\mathrm{new}}_2(a_2) \hat{\mathcal{O}}_3(0)\rangle$  is beyond the scope of the current paper, so we  temporarily ignore this mixing. Then we only compute the old correlator $\langle \hat{\mathcal{O}}^{\circ}_1(a_1)\hat{\mathcal{O}}^{\circ}_2(a_2) \hat{\mathcal{O}}_3(0)\rangle$ even for the special extremal case~\footnote{Strictly speaking, the results for three point functions in section~\ref{Section:overlaps} is rigorous  only for the special next-to-extremal case from the viewpoint of ABJM theory. Nevertheless,  the result on the integrability and the overlaps involving $|\mathcal{B}_0^{\mathrm{even}}\rangle$ and $|\mathcal{B}_{2L_3}^{\mathrm{even}}\rangle$ is still meaningful and helpful from the spin chain point of view. } .Then with $l_{31|2}=0$, we have,
	\begin{equation}
	\langle \hat{\mathcal{O}}_1^\circ(a_1) \hat{\mathcal{O}}_2^\circ(a_2) \hat{\mathcal{O}}_3(0)\rangle  =\frac{(-1)^{\frac{l_{12|3}}{2}}L_1L_2L_3\lambda^{\sum_{i=1}^3L_i}}{N|a_1|^{l_{31|2}} |a_2|^{l_{23|1}}} C_{I_1\cdots I_{L_3}}^{J_1\cdots J_{L_3}} 
	(\bar{n}_2)^{I_1}(n_2)_{J_1}\cdots 
	(\bar{n}_2)^{I_{L_3}}(n_2)_{J_{L_3}}\,,
\end{equation}
	and the result for the case with $l_{31|2}=2L_3$ is obtained by replacing $n_2, \bar{n}_2$ in this expression by $n_1, \bar{n}_1$.
	
	When $l_{pq|r}$'s are odd, Wick contractions lead to 
\begin{eqnarray}
			\label{WickOdd} && \langle \hat{\mathcal{O}}_1^\circ(a_1) \hat{\mathcal{O}}_2^\circ(a_2) \hat{\mathcal{O}}_3(0)\rangle  =\frac{(-1)^{\frac{l_{12|3}-1}{2}}\mathrm{sgn}(a_{12})L_1L_2\lambda^{\sum_{i=1}^3L_i}}{N|a_1|^{l_{31|2}} |a_2|^{l_{23|1}}} C_{I_1\cdots I_{L_3}}^{J_1\cdots J_{L_3}}\nonumber\\
		&\times&
		\sum_{1\leq s, s^\prime \leq L,\, s^\prime-s=\frac{l_{31|2}-1}{2}}
		\left((\bar{n}_2)^{I_1}(n_2)_{J_1}\cdots (n_2)_{J_{s-1}}(\bar{n}_1)^{I_s}\cdots (\bar{n}_1)^{I_{s^\prime}}(n_2)_{J_{s^\prime}}\cdots (n_2)_{J_{L_3}} \right.\nonumber \\
		&-& \left. (\bar{n}_2)^{I_1}(n_2)_{J_1}\cdots (\bar{n}_2)^{I_s} (n_1)_{J_s}\cdots (n_1)_{J_{s^\prime}}(\bar{n}_2)^{I_{s^\prime+1}}\cdots (n_2)_{J_{L_3}}\right)
		\,. 
\end{eqnarray}
	There are two classes of Wick contractions between $\hat{\mathcal{O}}^\circ_1$ and $\hat{\mathcal{O}}^\circ_2$. One leads to  the factor $d_{12}^{(l_{12|3}+1)/2} d_{21}^{(l_{12|3}-1)/2}$ and the other leads to a factor with $d_{12}$ and $d_{21}$ exchanged. This distinction results in the relative minus sign observed in~\eqref{WickOdd}.
	
	We demand that the non-BPS operator $\hat{\mathcal{O}}_3$ be an eigenstate of the planar two-loop dilatation operator within the spin chain framework. Under this condition, this operator is mapped to the following state
	\be |\hat{\mathcal{O}}_3\rangle= C_{I_1\cdots I_{L_3}}^{J_1\cdots J_{L_3}}| I_1, \bar{J}_1, \cdots, I_{L_3}, \bar{J}_{L_3}\rangle\,, \ee
	of the $SU(4)$ alternating  spin chain. Generically this state can be parametrized by the solution $\mathbf{u}, \mathbf{v}, \mathbf{w}$ of the Bethe ansatz equations~\cite{Minahan:2008hf,Bak:2008cp}, 
	\be |\hat{\mathcal{O}}_3\rangle=|\mathbf{u}, \mathbf{w}, \mathbf{v}\rangle\,. \ee
	The three point function $\langle \hat{\mathcal{O}}_1^\circ(a_1) \hat{\mathcal{O}}_2^\circ(a_2) \hat{\mathcal{O}}_3(a_3)\rangle$
	can be written as the overlap between a boundary state and the above Bethe state
	$|\mathbf{u}, \mathbf{w}, \mathbf{v}\rangle$.
	When $l_{pq|r}$'s are even and $2\leq l_{31|2}\leq 2L_3-2$, we define $\langle \mathcal{B}^{\mathrm{even}}_{l_{31|2}}|$ as follows 
	\begin{equation}
		\begin{aligned}
			\langle \mathcal{B}_{l_{31|2}}^{\mathrm{even}}|
			&&:=\sum_{1\leq s, s^\prime \leq L,\, s^\prime-s=\frac{l_{31|2}}{2}-1}
			\left((\bar{n}_2)^{I_1}(n_2)_{J_1}\cdots (n_2)_{J_{s-1}}(\bar{n}_1)^{I_s}\cdots (n_1)_{J_{s^\prime}}(\bar{n}_2)^{I_{s^\prime+1}}\cdots (n_2)_{J_{L_3}} \right.\nonumber \\
			&&+ \left. (\bar{n}_2)^{I_1}(n_2)_{J_1}\cdots (\bar{n}_2)^{I_s} (n_1)_{J_s}\cdots (\bar{n}_1)^{I_{s^\prime+1}}(n_2)_{J_{s^\prime+1}}\cdots (n_2)_{J_{L_3}}\right)\langle I_1, J_1, \cdots, I_{L_3}, J_{L_3}|
			\,.
		\end{aligned}
	\end{equation}
 Then that the above three-point function is proportional to the overlap between the boundary state and the Bethe state,
		\begin{equation}
		\langle \hat{\mathcal{O}}_1^\circ(a_1) \hat{\mathcal{O}}_2^\circ(a_2) \hat{\mathcal{O}}_3(0)\rangle 
		=\frac{(-1)^{\frac{l_{12|3}}{2}}L_1L_2\lambda^{\sum_{i=1}^3L_i}}{N|a_1|^{l_{31|2}} |a_2|^{l_{23|1}}} \langle
\mathcal{B}_{l_{31|2}}^{\mathrm{even}}|\mathbf{u}, \mathbf{w}, \mathbf{v}\rangle\,.
		\label{BoundaryEven}
	\end{equation}
	
	To clarify the structure of $\langle\mathcal{B}_{l_{31|2}}^{\mathrm{even}}|$, we consider the operator $U_{\mathrm{even}}$~\cite{Jiang:2023cdm}, $U_{\mathrm{odd}}$, which shift all even or odd sites to the left by two units, respectively~\footnote{We define $L$ to be $L_3$.}, 
	\begin{align}
		&U_{\mathrm{even}}|I_1, J_1, I_2, J_2\cdots, I_{L-1}, J_{L-1}, I_L, J_L\rangle\nonumber \\  & =|I_1, J_2, I_2, J_3, \cdots, I_{L-1}, J_L, I_L, J_1\rangle\,, \\
		& U_{\mathrm{odd}}|I_1, J_1, I_2, J_2\cdots, I_{L-1}, J_{L-1}, I_L, J_L\rangle\nonumber \\ & =|I_2, J_1, I_3, J_2, \cdots, I_L, J_{L-1}, I_1, J_L\rangle\,.
	\end{align}
	
	We further define, for $m=1, \cdots, L$,
	\begin{align}
		&\langle\bar{n}_1@\{1, 2,\cdots, m\},  n_1@\{1, 2, \cdots, m\}|=\nonumber \\ & (\bar{n}_1)^{I_1}(n_1)_{J_1}
		\cdots (\bar{n}_1)^{I_m}(n_1)_{J_m}(\bar{n}_2)^{I_{m+1}}(n_2)_{J_{m+1}}\nonumber \\ & \cdots (\bar{n}_2)^{I_1}(n_2)_{J_L} 
		\langle I_1, J_1, \cdots, I_L, J_L|\,.
	\end{align}
	Here the number labels the sites on the spin chain.
	Then we can express $\langle \mathcal{B}_{l_{23|1}}^{\mathrm{even}}|$
	as
	\be \langle \mathcal{B}_{l_{23|1}}^{\mathrm{even}}|=
	\langle \mathcal{B}_{l_{23|1}}^{\mathrm{even}, \,a}|+\langle \mathcal{B}_{l_{23|1}}^{\mathrm{even},\, b}|\,\ee
	
	with
\begin{equation}
		\begin{aligned}\langle\mathcal{B}_{l}^{\mathrm{even}, \,a}|
		&=\sum_{j=0}^{L-1}  \langle\bar{n}_1@\{1, 2,\cdots, l/2\},  {n}_1@\{1, 2, \cdots, l/2\}|(U_{\mathrm{even}}U_{\mathrm{odd}})^j\,,\\
		\langle\mathcal{B}_{l}^{\mathrm{even}, \,b}|&=\langle\mathcal{B}_{l}^{\mathrm{even}, \,a}|U_{\mathrm{even}}\,. 
	\end{aligned}
\end{equation}
	The corresponding  boundary states for odd $l_{pq|r}$, as well as the cases with $l_{31|2}=0$ or $2L_3$, are presented in the Appendix~\ref{appendix1}.

\section{ Integrable boundary states from two BPS operators}\label{Section:integrability}
It is interesting to investigate whether the boundary states associated with two BPS single-trace operators, namely $\langle \mathcal{B}^{\mathrm{even}}_l|$ or $\langle \mathcal{B}^{\mathrm{odd}}_l| $, are integrable in the sense defined in~\cite{Piroli:2017sei, Gombor:2020kgu}.
It was demonstrated in~\cite{Yang:2022dlk} that the boundary state $\langle\mathcal{B}^{\mathrm{even}}_0|$ and 
$\langle\mathcal{B}^{\mathrm{even}}_{2L}|$
satisfy the following twisted integrable condition~\cite{Gombor:2020kgu}, 
\be\label{integarble0} \tau(\lambda)|\mathcal{B}^{\mathrm{even}}_l \rangle=
\tau(-\lambda-2)|\mathcal{B}^{\mathrm{even}}_l \rangle,\ee 
where $l=0$ or $2L$ and $\tau(\lambda)$ is one of two transfer matrices defined in \eqref{transfer}.

Now we will prove that $\langle \mathcal{B}^{\mathrm odd}_1|$ is integrable. This boundary state arises from the special next-to-extremal three-point functions. In fact, we shall verify a stronger claim: both $\langle \mathcal{B}^{{\mathrm odd},\, a}_1|$ and $\langle \mathcal{B}^{{\mathrm odd},\, b}_1|$
are integrable, respectively. We will provide details for the case of  $\langle \mathcal{B}^{{\mathrm odd}, \, b}_1|$, as the analysis for $\langle \mathcal{B}^{{\mathrm odd}, \, a}_1|$ is entirely analogous. %We just focus on  $|\langle \mathcal{B}^{\mathrm{odd}, \, a}_1|$.
Recall that \bea |\mathcal{B}^{{\mathrm odd},\, b}_1\rangle&=&\sum_{j=0}^{L-1}(U_{\mathrm{even}}U_{\mathrm{odd}})^j|n_1@1\rangle\nonumber\\
&=&\sum_{j=1}^L |n_1@j\rangle\,.
\eea
Here, we simply denote $|n_1 @\{1\}\rangle$ by $|n_1 @ 1\rangle$, and $|n_1@j \rangle$ represents 
\begin{equation}(n_2)^{I_1}(\bar{n}_2)_{J_1}\cdots (\bar{n}_2)_{J_{j-1}}(n_1)^{I_j}(\bar{n}_2)_{J_j}\cdots (n_2)^{I_L}(\bar{n}_2)_{J_L}|I_1,J_1,\cdots I_L,J_L\rangle\,.\end{equation}

Let us decompose $\tau(\lambda)$ as
\begin{equation}
	\tau(\lambda)=\sum_{m=0}^{L}\sum_{n=0}^{L} \lambda^{L-m}(-\lambda-2)^{L-n}{\mathcal O}_{m, n}\,,
\end{equation}
with the condition that,   for each term of ${\mathcal O}_{m, n}$,~\footnote{To distinguish  the operators from  the decomposition of $\tau(\lambda)$ and the single-trace operator in ABJM theory, we use $\mathcal{O}$ without hat to denote the former.} there are exactly $m$  $\mathbb{P}$'s and $n$ $\mathbb{K}$'s inside the trace. Under this condition, the decomposition is unique.

Then we have
\begin{eqnarray}
	% \nonumber % Remove numbering (before each equation)
	\tau(-\lambda-2)&=&\sum_{m=0}^{L}\sum_{n=0}^{L} (-\lambda-2)^{L-m}\lambda^{L-n}{\mathcal O}_{m, n} \nonumber\\
	&=&\sum_{m=0}^{L}\sum_{n=0}^{L} \lambda^{L-m}(-\lambda-2)^{L-n}{\mathcal O}_{n, m}\,.
\end{eqnarray}

As in~\cite{Yang:2022dlk}, the twisted integrable condition 
\begin{equation}\label{IntegrableStates}
	\tau(\lambda)|\mathcal{B}_1^{\mathrm{odd}, \, b}\rangle=\tau(-\lambda-2)|\mathcal{B}_1^{\mathrm{odd},\, b}\rangle\,,
\end{equation}is satisfied when the boundary state $||\mathcal{B}_1^{\mathrm{odd}, \, b}\rangle$ satisfies
\begin{equation}\label{criterion}
	{\mathcal O}_{m, n}|\mathcal{B}_1^{\mathrm{odd}, \, b}\rangle={\mathcal{O}}_{n, m}|\mathcal{B}_1^{\mathrm{odd,}\, b}\rangle.
\end{equation}
for any $0\leq m, n\leq L$. The above equations provide a sufficient condition for $|\mathcal{B}^{\mathrm{odd},\, b}_1\rangle$ to satisfy the twisted integrable condition.
Although terms in  $\mathcal{O}_{m, n}$'s are  non-local operators for $m+n\geq 2$, these sufficient condition is often easier to verify rigorously for certain boundary states compared to the criteria involving higher conserved charges obtained from \eqref{IntegrableStates}. %, as demonstrating in~\cite{Yang:2022dlk} and the proof here.

As a first step, it is evident that these equations~\eqref{criterion} hold when $m=n$ including the special case with $m=n=0$.

It is also straightforward to observe that, 
\bea {\mathcal O}_{1, 0}&=&\sum_{i=1}^L {\mathrm tr}_0(\mathbb{P}_{0i})=L \mathbb{I}\,,\\
{\mathcal O}_{0, 1}&=&\sum_{j=1}^L {\mathrm tr}_0(\mathbb{K}_{0\bar{j}})=L \mathbb{I}\,,\eea
so \be {\mathcal O}_{1, 0}|\mathcal{B}^{\mathrm{odd}, \, b}_1\rangle={\mathcal O}_{0, 1}|\mathcal{B}^{\mathrm{odd}, \, b}_1\rangle. \ee
Here $a$ denotes the auxiliary space index.

For $m>1$, we have
\begin{eqnarray}
	{\mathcal O}_{m, 0}&=&\sum_{i_1<\cdots <i_m} {\mathrm tr}_a({\mathbb{P}}_{ai_1}\cdots {\mathbb P}_{ai_m})\nonumber\\
	&=& \sum_{i_1<\cdots <i_m}  {\mathbb{P}}_{i_m, i_1}{\mathbb{P}}_{i_m, i_2}\cdots {\mathbb P}_{i_m, i_{m-1}}\,.
\end{eqnarray}
Let us denote $ {\mathbb{P}}_{i_m, i_1}{\mathbb{P}}_{i_m, i_2}\cdots {\mathbb P}_{i_m, i_{m-1}}$ as ${\mathcal O}^{\mathbb P}_{i_1 \cdots i_m}$.
we can find that
\begin{align}
	&  {\mathcal O}^{\mathbb P}_{i_1, i_2, \cdots, i_m}|I_{i_1}, I_{i_2}, \cdots, I_{i_{m-1}}, I_{i_m}\rangle=\nonumber\\ &|I_{i_2}, I_{i_3}, \cdots, I_{i_m},  I_{i_1}\rangle\,.
\end{align}
Here, we omit the labels in the ket state that are not acted upon by the operator $ {\mathcal O}^{\mathbb P}_{i_1, i_2, \cdots, i_m}$. Thus, the action of this operator results in a cyclic permutation among the indices $i_1, i_2, \cdots,i_m$. From this, we determine the action of this operator on $| n_1 @ i\rangle$ and further on $| \mathcal{B}_1^{\mathrm{odd}}\rangle$. Specifically, when $i\neq i_1, \cdots, i_m$, we have
\begin{equation}
	{\mathcal O}^{\mathbb P}_{i_1 \cdots i_m}|n_1@i\rangle=|n_1@i\rangle\,.
\end{equation}
On the other hand,
\begin{equation}
	{\mathcal O}^{\mathbb P}_{i_1 \cdots i_m}|n_1@i_k\rangle=|n_1@i_{k-1}\rangle\,,
\end{equation}
where the index $k$ is understood as $\mathrm{mod}\, m$.
From these results, we have 
\begin{equation}
	{\mathcal O}^{\mathbb P}_{i_1 \cdots i_m}\left(\sum_{k=1}^{m}|n_1@i_k\rangle\right)=\sum_{k=1}^{m}|n_1@i_k\rangle\,.
\end{equation}
Then
\begin{equation}
	{\mathcal O}^{\mathbb P}_{i_1 \cdots i_m}|\mathcal{B}_1^{{\mathrm{odd}}, \, b}\rangle=|\mathcal{B}_1^{{\mathrm{odd}}, \, b}\rangle\,,
\end{equation}
\begin{equation}
	{\mathcal O}_{m, 0}|\mathcal{B}_1^{{\mathrm{odd}}, \, b}\rangle=\binom{L}{m}|\mathcal{B}_1^{{\mathrm{odd}}, \, b}\rangle\,,  \end{equation}
where $\binom{L}{m}$ is the binomial coefficient.

We now turn to the operator $\mathcal{O}_{0, m}$,
\begin{eqnarray}
	% \nonumber % Remove numbering (before each equation)
	{\mathcal O}_{0, m}&=&\sum_{j_1<\cdots <j_m} {\mathrm tr}_a({\mathbb{K}}_{a\bar{j}_1}\cdots {\mathbb K}_{a\bar{j}_m})\\
	&=& \sum_{j_1<\cdots <j_m}  {\mathbb{P}}_{\bar{j}_m, \bar{j}_{m-1}}{\mathbb{P}}_{\bar{j}_m, \bar{j}_{m-2}}\cdots {\mathbb P}_{\bar{j}_m, \bar{j}_{1}}\,.
\end{eqnarray}
Since the above permutations only act on  the even sites, their action will not change $|n_1@i\rangle$. Then the action of $\mathcal{O}_{0, m}$ is
\begin{eqnarray}
	{\mathcal O}_{0, m}|n_1@i\rangle&=&\binom{L}{m}|n_1@i\rangle\,,  \\
	{\mathcal O}_{0, m}|\mathcal{B}_1^{{\mathrm{odd}}, \, b}\rangle&=&\binom{L}{m}|\mathcal{B}_1^{{\mathrm{odd}},\, b}\rangle\,.
\end{eqnarray}
Thus
\begin{equation}
	{\mathcal O}_{m, 0}|\mathcal{B}_1^{{\mathrm{odd}}, \, b}\rangle=\binom{L}{m}|\Psi_1\rangle={\mathcal O}_{0, m}|\mathcal{B}_1^{{\mathrm{odd}},\, b}\rangle\,.
\end{equation}

The proof of \eqref{criterion} for general $m,n\neq0$ is deferred to the Appendix~\ref{appendixA}. 
With this, we finish the proof of \eqref{criterion}.

The proof that $\langle\mathcal{B}_{2L_3-1}^{\mathrm{odd}}|$ is integrable follows similarly, since  this  bra state can be obtained by  exchanging $(n_1, \bar{n}_1)$ and $(n_2, \bar{n}_2)$ in $\langle \mathcal{B}_1^{\mathrm{odd}}|$  as mentioned before. Generally  $\langle\mathcal{B}_{l_{31|2}} |$ and $\langle\mathcal{B}_{2L-l_{31|2}}|$  are related by exchanging the roles of $\hat{\mathcal{O}}_1$ and $\hat{\mathcal{O}}_2$, regardless of $l_{31|2}$ being even or odd.

Notice that in the above proof, we do not rely on the BPS conditions, $\bar{n}_1\cdot n_1=\bar{n}_2\cdot n_2=0$, letting alone the twisted-translated frame. Therefore, the proof demonstrates that the more general boundary states
$\langle \mathcal{B}^{{\mathrm{odd}, \, b}}_1|$ and $\langle \mathcal{B}^{{\mathrm{odd}, \, a}}_1|$ without any constraints on $n_i$'s and $\bar{n}_i$'s are integrable. This also implies that the states $\langle\mathcal{B}^{{\mathrm{odd}}, \, a}_{2L-1}|$ and  $\langle\mathcal{B}^{{\mathrm{odd}}, \, b}_{2L-1}|$ are integrable  even without the BPS constraints.

A set of  unpaired Bethe roots with $L=3, N_{\mathbf{u}}=N_{\mathbf{w}}=1, N_{\mathbf{v}}=2$,
\be
u_1=0.866025,\,\quad w_1= 0.866025,\,\quad v_1=-0.198072,\,\quad v_2= 0.631084\,,
\ee
was found numerically  in~\cite{Jiang:2023cdm}. The corresponding Bethe state can be constructed using the coordinated Bethe ansatz in~\cite{Yang:2021hrl}. We found that the  overlaps between this Bethe state and some boundary states $|\mathcal{B}_{l_{31|2}}\rangle$
  after a general $SU(4)$ rotation~\footnote{Without such rotation, the overlaps  trivially vanish because the selection rule $N_{\mathbf{u}}=N_{\mathbf{v}}=N_{\mathbf{w}}$, given in the next section, is not satisfied by this set of Bethe roots.  }  are non-zero when  $l_{31|2}\neq 0, 1, 2L_3-1, 2L_3$. This also applies for such boundary states from the three-point functions in the twisted-translated frame.
Based on this, we conjecture that  the boundary state is integrable only when  the three-point function is special (next-to-)extremal.

\section{The overlaps}\label{Section:overlaps}
Due to the construction of the twisted-translated frame, the scalar fields in $\hat{\mathcal{O}}_1^\circ$ and $\hat{\mathcal{O}}_2^\circ$ can only be one of $Y^1, Y^4, Y^\dagger_1, Y^\dagger_4$. This restriction leads to the following selection rule,
\be N_{\mathbf{u}}=N_{\mathbf{w}}=N_{\mathbf{v}}\,, \ee
for the overlap $\langle \mathcal{B}_{l_{31|2}}|\mathbf{u}, \mathbf{v}, \mathbf{w}\rangle$ to be nonzero, here $\mathcal{B}$ is either $\mathcal{B}^{\mathrm{even}}$
or $\mathcal{B}^{\mathrm{odd}}$, depending on whether $l_{31|2}$ is even or odd, and $N_{\mathbf{u}}, N_{\mathbf{w}}, N_{\mathbf{v}}$ are numbers of three types of Bethe roots ${\mathbf{u}}, {\mathbf{w}}, {\mathbf{v}}$.

One consequence of the result~\eqref{constraint} from symmetries %and the definitions of the boundary states~\eqref{BoundaryEven}, \eqref{BoundaryOdd} 
is an upper bound for $N_{\mathbf{u}}$, 
\be\label{UpperBound} N_{\mathbf{u}}\leq \mathrm{min}(l_{31|2}, l_{23|1})\,. \ee
To see this, first we notice that $\Delta_3-J_3=N_{\mathbf{u}}$ where $\Delta_3$ and $J_3$ are the tree-level conformal dimension and the $U(1)$ R-charge of the operator $\hat{\mathcal{O}}_3$, respectively. 
Then~\eqref{constraint} gives, 
\be\frac{\langle \hat{\mathcal{O}}_1^\circ(a_1) \hat{\mathcal{O}}_2^\circ(a_2) \hat{\mathcal{O}}_3(0)\rangle}{\sqrt{\mathcal{N}_{\hat{\mathcal{O}}^\circ_1}\mathcal{N}_{\hat{\mathcal{O}}^\circ_2}\mathcal{N}_{\hat{\mathcal{O}}_3}}}=\mathcal{C}_{123}\left(\frac{a_{12}}{a_{2}a_{1}}\right)^{N_{\mathbf{u}}} \,.\ee
Now from 
\eqref{BoundaryEven} and \eqref{BoundaryOdd}, we get 
	\bea \langle \mathcal{B}_{l_{31|2}}|\mathbf{u}, \mathbf{w}, \mathbf{v} \rangle &\propto& \mathcal{C}_{123} a_1^{l_{31|2}} a_2^{l_{23|1}} \frac{a_{12}^{N_{\mathbf{u}}}}{a_1^{N_{\mathbf{u}}}a_2^{N_{\mathbf{u}}}}\nonumber\\
&\propto&\mathcal{C}_{123} \sum_{j=0}^{N_{\mathbf{u}}} (-1)^{N_\mathbf{u}-j}\binom{N_\mathbf{u}}{j}  a_1^{l_{31|2}-N_{\mathbf{u}}+j}a_{2}^{l_{23|1}-j}\,,\eea
where in both steps we omit the possible minus signs and other factors independent of $a_i$'s. And here the binomial expansion of $a_{12}^{N_{\mathbf{u}}}=(a_1-a_2)^{N_{\mathbf{u}}}$ has been used.
Since each term in $\langle \mathcal{B}_{l_{31|2}}|\mathbf{u}, \mathbf{w}, \mathbf{v} \rangle$ should have non-negative exponents of $a_1$ and $a_2$, we obtain
\be l_{31|2}-N_{\mathbf{u}}+j\geq 0\,, \, 
l_{23|1}-j\geq 0\,,\ee
for $j=0, 1, \cdots, N_{\mathbf{u}}$. This leads to the upper bound given in \eqref{UpperBound}. This bound significantly simplifies the computations of the overlap between the integrable boundary state from two $1/3$-BPS operators and the Bethe state.

The only proven integrable boundary states are those with $l_{31|2}=0, 1, 2L_3-1, 2L_3$. As discussed before, we only need to compute the correlator with $l_{31|2}=0$ or $1$. The result for the other two case can be obtained by exchanging $(n_1, \bar{n}_1)$ and $(n_2, \bar{n}_2)$. 
Let us first consider the special extremal case with $l_{31|2}=0$. Now the above upper bound gives $N_{\mathbf{u}}=0$, and the only Bethe state in this sector is just the vacuum state, 
\be | \emptyset, \emptyset, \emptyset\rangle=|1\bar{4}\rangle^{\otimes^L}\,, \ee
and now the three-point function is a correlator of three BPS operators.

It is easy to obtain that 
\bea\langle \mathcal{B}_0^{\mathrm{even}}|\emptyset, \emptyset, \emptyset\rangle&=&L\left((\bar{n}_2)^1(n_2)_4\right)^L\nonumber\\
&=&
(-1)^{L}La_2^{2L}\,. \eea
From this result,  we obtain the OPE coefficient ${\mathcal C}_{123}$  in~\eqref{constraint} as 
\be {\mathcal C}_{123}=\frac{(-1)^{L_2}\sqrt{L_1L_2L_3}}{N}\,, \ee
where \eqref{BoundaryEven}, the condition  $l_{31|2}=0$ and the normalization factors in~\eqref{normBPS} for all three $1/3$-BPS operators have been used.

Now we turn to the special next-to-extremal case with $l_{31|2}=1$. According to the upper  bound in~\eqref{UpperBound}, we get $N_{\mathbf{u}}\leq 1$. 
Thus, we need to examine two sub-cases with $N_{\mathbf{u}}=0, 1$. 
In the sub-case
with $N_{\mathbf{u}}=0$, we have 
\begin{equation}
	\begin{aligned}
			\langle\mathcal{B}^{\mathrm{odd},\, a}_1| \emptyset, \emptyset, \emptyset \rangle=\sum_{j=1}^L\langle\bar{n}_1@j | \emptyset, \emptyset, \emptyset \rangle=(-1)^LL a_1 a_2^{2L-1}\,, \nonumber \\
		\langle\mathcal{B}^{\mathrm{odd},\, b}_1| \emptyset, \emptyset, \emptyset \rangle=\sum_{j=1}^L\langle n_1@j | \emptyset, \emptyset, \emptyset \rangle=(-1)^LL a_1 a_2^{2L-1}\,.
	\end{aligned}
\end{equation}
So 
\bea\langle\mathcal{B}^{\mathrm{odd}}_1| \emptyset, \emptyset, \emptyset \rangle&=&\langle\mathcal{B}^{\mathrm{odd},\, a}_1| \emptyset, \emptyset, \emptyset \rangle-\langle\mathcal{B}^{\mathrm{odd}, \, b}_1| \emptyset, \emptyset, \emptyset \rangle\nonumber\\&=&0\,,\eea
\be \mathcal{C}_{123}=0\,.\ee

Now we turn to the sub-case with  $N_{\mathbf{u}}=N_{\mathbf{w}}=N_{\mathbf{v}}=1$. Now the Bethe roots $u, v, w$ should satisfy the Bethe ansatz equations \eqref{Bethe equations} and zero momentum condition \eqref{zero momentum}, and the solutions of these equations are~\cite{Bak:2008cp}, 
\be\label{betheroots} u=-v=\frac{1}{2}\cot{\frac{k\pi}{L+1}}\,, \,\, w=0\,,\ee
where $ k=1, \cdots, L$.
The Bethe state $|\{u\}, \{w\}, \{v\}  \rangle$ can be constructed using the nested coordinate Bethe ansatz in~\cite{Yang:2021hrl}. This Bethe state can be written as 
\begin{align}
	&   |\{u\}, \{w\}, \{v\}  \rangle\nonumber\\
	=&\sum_{n=1}^L\Phi(\overset{\overset{w}{u, v}}{\bullet_n})\,|\overset{\overset{w}{u, v}}{\bullet_n}\rangle+\sum_{n=1}^L\Phi(\overset{\overset{w}{u, v}}{\circ_n})\,|\overset{\overset{w}{u, v}}{\circ_n}\rangle+\cdots\,.\label{betheState}
\end{align}
Some remarks are in order here. First, only the terms which are explicitly displayed contributes to the overlap
$\langle\mathcal{B}_1^{\mathrm{odd}} | \{u\}, \{w\}, \{v\}\rangle$. Second, $|\overset{\overset{w}{u, v}}{\bullet_n}\rangle$  and $|\overset{\overset{w}{u, v}}{\circ_n}\rangle$ stand for 
\begin{align}
	&|\bullet_1, \circ_1, \cdots, \overset{\overset{w}{u, v}}{\bullet_n}, \circ_n, \cdots, \bullet_L, \circ_L\rangle=\nonumber\\ &|1 \bar{4} 1 \bar{4} \cdots 4 \bar{4} \cdots 1 \bar{4}\rangle\,, 
\end{align}
and 
\begin{align}
	&|\bullet_1, \circ_1, \cdots, \bullet_n, \overset{\overset{w}{u, v}}{\circ_n}, \cdots, \bullet_L, \circ_L\rangle=\nonumber \\ &|1 \bar{4} 1 \bar{4} \cdots 1 \bar{1} \cdots 1 \bar{4}\rangle\,, 
\end{align}
respectively.

The displayed coefficients in~\eqref{betheState} are
\bea \Phi(\overset{\overset{w}{u, v}}{\bullet_n})&=&\left(\frac{u+\frac{\ii}{2}}{u-\frac{\ii}{2}}\frac{v+\frac{\ii}{2}}{v-\frac{\ii}{2}}\right)^n\frac{-(v-\frac{\ii}{2})}{(w-u-\frac{\ii}{2})(w-v-\frac{\ii}{2})}\nonumber\\
&=&\frac{2}{\ii-2u}\,,\nonumber \\
\Phi(\overset{\overset{w}{u, v}}{\circ_n})&=&\left(\frac{u+\frac{\ii}{2}}{u-\frac{\ii}{2}}\frac{v+\frac{\ii}{2}}{v-\frac{\ii}{2}}\right)^n\frac{u+\frac{\ii}{2}}{(w-u-\frac{\ii}{2})(w-v-\frac{\ii}{2})}\nonumber\\
&=&\frac{2}{\ii-2u}\,,\eea
where $u=-v, w=0$ has been used. Then 
\bea  \langle \mathcal{B}^{\mathrm{odd}}_1| \{u\}, \{0\}, \{-u\}\rangle
&=& \frac{2}{\ii-2u}\sum_{j=1}^L\sum_{n=1}^L \left( \langle \bar{n}_1 @j |\overset{\overset{w}{u, v}}{\bullet_n} \rangle+\langle \bar{n}_1 @j |\overset{\overset{w}{u, v}}{\circ_n} \rangle-\langle n_1 @j |\overset{\overset{w}{u, v}}{\bullet_n} \rangle
\right. \nonumber \\
&-&  \left. \langle n_1 @j |\overset{\overset{w}{u, v}}{\circ_n} \rangle\right)\eea
After some computations, it is not difficult to obtain
\bea  &&\langle\mathcal{B}_1^{\mathrm{odd}} | \{u\}, \{0\}, \{-u\}\rangle\nonumber\\
&=&{{4 (-1)^L La_2^{2L-2}a_{12}}\over {\ii-2u}}\nonumber\\
&=&
4(-1)^{L+1} La_2^{2L-2}a_{12}\sin\frac{k\pi}{L+1}\exp\frac{k\pi\ii}{L+1}\,, \eea
where \eqref{betheroots} has been used. By applying this result along with \eqref{BoundaryOdd} and $l_{31|2}=1$,

we can obtain the result of the  three point function 
\bea &&\langle \hat{\mathcal{O}}_1^\circ(a_1)\hat{\mathcal{O}}_2^\circ(a_2) \hat{\mathcal{O}}_3(0) \rangle\nonumber\\
&=&\frac{4(-1)^{L_2+1}}{N} \mathrm{sgn}(a_1 a_2 a_{12})\left(\prod_{j=1}^3 L_j\right) \,\lambda^{\sum_{k=1}^3 L_k}\,\frac{a_{12}}{a_1 a_2}\sin\frac{k\pi}{L_3+1}
\exp\frac{k\pi \ii  }{L_3+1}\,. \eea

At planar tree-level order, we have, 
\begin{align}
	\mathcal{N}_{\hat{\mathcal{O}}^\circ_{L_j}}&=L_j \lambda^{2L_j}\,,\,\, j=1\,, 2\,,\\
	{\mathcal{N}_{\hat{\mathcal{O}}_3}}&=L_3 \lambda^{2L_3}\langle \{u\}, \{0\}, \{-u\}|\{u\}, \{0\}, \{-u\}  \rangle\,,
\end{align}
with the norm of $|\{u\}, \{0\}, \{-u\}\rangle$  being,
\bea && \langle\{u\}, \{0\}, \{-u\}|\{u\}, \{0\}, \{-u\}\rangle\nonumber\\
&=&\big{|}(u^2+\frac{1}{4})^2 \,\mathrm{det}G\big{|} \nonumber\\
&=&8L_3(L_3+1)\sin^2\frac{k\pi}{L_3+1}\,, \eea
where a formula for the norm in~\cite{Yang:2021hrl} has been used and $\mathrm{det}G$ is the Gaudin determinant.
Finally we get the structure constant $\mathcal{C}_{123}$ in \eqref{constraint} as 
\be \mathcal{C}_{123}=\frac{(-1)^{L_2+1}\mathrm{sgn}(a_1 a_2 a_{12})\sqrt{2L_1L_2L_3}\exp\frac{2k\pi \ii}{L_3+1}}{N \sqrt{L_3+1}}\,.\ee

\section{Outlook}\label{outlook}
In this paper, we stuied three-point correlators of single-trace operators  in ABJM theory using integrable boundary states. We analyzed correlation functions of two $1/3$-BPS and one non-BPS operator, observing that the three-point function is proportional to the overlap between a boundary state and a Bethe state. The resulting boundary state is integrable when the correlator  is special (next-to-)extremal. For the special next-to-extremal case,
We computed these three-point functions using symmetry constraints and the coordinate Bethe ansatz.

For the special next-to-extremal case with $l_{31|2}=1$, the corresponding integrable boundary states are a sum of relatively simple states $|n_1@j\rangle$ and $|\bar{n}_1 @j\rangle$ which in turn are similar to special domain wall states which appeared in the study of integrable quantum circuits~\cite{Hutsalyuk:2024saw}. Notice that these integrable boundary states and the integrable $S_L$-invariant states mentioned in the appendix~\ref{appendix3} are neither two-site product states nor matrix product states, unlike many previously studied integrable boundary states in AdS/CFT correspondence. This shows the power of the sufficient 
condition (criteria)  in the section~\ref{Section:integrability} (Appendix~\ref{appendix3}) on finding new type of integrable states.

It is interesting to revisit similar three-point functions in ${\mathcal N}=4$ SYM~\cite{Escobedo:2010xs} to identify the conditions under which the obtained boundary states are integrable. We are also interested in computing more general three-point functions in ABJM theory to aid the development of the hexagon program~\cite{Basso:2015zoa} for ABJM theory.

\section*{Acknowledgments}
We would like to thank Yunfeng~Jiang for collaboration at  early stages of this project, Kun~Hao, Jian-Xin~Lu,  Hong~Lu, Pujian~Mao, Zhao-Long~Wang and Jia-ju~Zhang for help discussions.
J.~W. would like to thank the Center for Theoretical Physics, Sichuan University and Shing-Tung Yau Center of Southeast University during \textit{International Workshop on Exact Methods in Quantum Field Theory and String Theory} for warm hospitality. J.~W.  would also like to thank Ningbo Lishe International Airport where part of the calculation was performed. 
This work is partly supported  by the National Natural Science Foundation of China, Grant No.~12375066, 11935009, 12247103.

\appendix

\section{Criteria for integrable boundary states in the $SU(N)$  spin chain}\label{appendix3}
We can establish  criteria for integrable boundary states for the closed $SU(N)$ spin chain with
the spin on each site being in the fundamental representation of  $SU(N)$. 
The Hamiltonian on this chain is~\cite{1970JETPL..12..225U, Sutherland:1975vr}
\bea H&=&\sum_{j=1}^L \mathbb{P}_{j, j+1}\,.\eea
%\bea H&=&\sum_{j=1}^L h_{j, j+1}\,, \\
%h_{j, j+1}&=&\mathbb{I}-\mathbb{P}_{j, j+1}\,.\eea

 Here the permutation operator $\mathbb{P}$ is defined as 
 \bea \mathbb{P}_{lm}|i\rangle_l\otimes |j\rangle_m&=&|j\rangle_l\otimes |i\rangle_m\,, 
  \eea
 where $|i\rangle_l, i=1, \cdots, N$ is the set of orthonormal basis of the Hilbert space at the $l$-th site. 

The corresponding $R$-matrix  is 
\be R_{ab}(\lambda)=\lambda \mathbb{I}+\mathbb{P}_{ab}\,.\ee
This $R$-matrix satisfies the Yang-Baxter equation which  ensure that the Hamiltonian derived from it is integrable.
The corresponding transfer matrix is 
\be \tau(\lambda)=\mathrm{tr}_0(R_{01}(\lambda)R_{02}(\lambda)\cdots R_{0L}(\lambda))\,. \ee

We define the parity operator $\Pi$ via its action on the basis $|i_1, i_2, \cdots, i_{L-1}, i_L\rangle$, 
\be  \Pi|i_1, i_2, \cdots, i_{L-1}, i_L\rangle=|i_L, i_{L-1}, \cdots, i_2, i_1\rangle\,.\ee
A state  $|\mathcal{B}\rangle$ in the Hilbert space of this spin chain is called an integrable boundary state~\cite{Piroli:2017sei} if it satisfies
\be \Pi \tau(\lambda) \Pi |\mathcal{B}\rangle= \tau(\lambda) |\mathcal{B}\rangle\,.\label{integrableSU} \ee 
$\tau(\lambda)$ is a polynomial of $\lambda$  with degree $L$. 
Consider its expansion as 
\be \tau(\lambda)=\sum_{m=0}^L \lambda^{L-m}\mathcal{O}_m\,.\ee
Then there are $m$ $\mathbb{P}$'s in each term of $\mathcal{O}_m$. 
So the integrable condition~\eqref{integrableSU} is equivalent to the following condition
\be  \mathcal{O}_m|\mathcal{B}\rangle = \Pi \mathcal{O}_m \Pi |\mathcal{B}\rangle\,, m=0, 1, \cdots, L. \label{SUOm}\ee

When $m=0, 1$, we have $\mathcal{O}_m=\mathbb{I}$, the above condition is trivially satisfied. 
When $2\leq m \leq L $, we get
\bea \mathcal{O}_m&=& \sum_{i_1<i_2<\cdots <i_m}\mathrm{tr}_0(\mathbb{P}_{0i_1}\mathbb{P}_{0i_2}\cdots \mathbb{P}_{0i_m})
\nonumber\\
&=&\sum_{i_1<i_2<\cdots <i_m}\mathbb{P}_{i_m, i_1}\mathbb{P}_{i_m, i_2}\cdots \mathbb{P}_{i_m, i_{m-1}}\,.\eea
Then 
\be\Pi \mathcal{O}_m \Pi= \sum_{j_1>j_2>\cdots >j_m}\mathbb{P}_{j_m, j_1}\mathbb{P}_{j_m, j_2}\cdots \mathbb{P}_{j_m, j_{m-1}}\,.\ee
Then  \eqref{SUOm} leads to the conditions
\bea  \sum_{i_1<i_2<\cdots <i_m}\mathbb{P}_{i_m, i_1}\mathbb{P}_{i_m, i_2}\cdots \mathbb{P}_{i_m, i_{m-1}}|\mathcal{B}\rangle 
=\sum_{i_1>i_2>\cdots >i_m}\mathbb{P}_{i_m, i_1}\mathbb{P}_{i_m, i_2}\cdots \mathbb{P}_{i_m, i_{m-1}}|\mathcal{B}\rangle\,, \eea
for $m=2, 3, \cdots, L$. Note that the case for $m=2$ is automatically satisfied. The remaining $L-2$ equations form the criteria for a state in this spin chain to be integrable.

An arbitrary permutation $\sigma \in S_L$ acts naturally on the Hilbert space of the $SU(N)$ spin chain as follows:
\be\sigma|i_1, \cdots, i_L\rangle=|i_{\sigma(1)}, \cdots, i_{\sigma(L)}\rangle\,. \ee
One direct consequence of the above criteria is that $S_L$-invariant states are all integrable. Among these states, it would be interesting to identify those whose overlaps with Bethe states have concise expressions in terms of Bethe roots via Gaudin determinants. 

\section{Basic notations of $SU(4)$ alternating spin chain and integrable boundary states}~\label{Section:notations}
A  single-trace operator in the scalar section of ABJM theory  corresponds to a state of alternating spin chain where odd and even sites are in the fundamental and anti-fundamental representations of  $SU(4)$, respectively.
	The planar two-loop dilatation operators in the scalar sector of ABJM theory is mapped to  the integrable Hamiltonian \cite{Minahan:2008hf, Bak:2008cp},
	\begin{equation}
		H=\frac{\lambda^2}{2}\sum_{l=1}^{2L}\left(2-2\mathbb{P}_{l,l+2}+\mathbb{P}_{l,l+2}\mathbb{K}_{l,l+1}+\mathbb{K}_{l,l+1}\mathbb{P}_{l,l+2}\right)\,,
	\end{equation}
 with the permutation operator $\mathbb{P}$ and the trace operator $\mathbb{H}$ defined as 
 \bea \mathbb{P}_{lm}|a\rangle_l\otimes |b\rangle_m&=&|b\rangle_l\otimes |a\rangle_m\,,  \\
 \mathbb{K}_{lm}|a\rangle_l\otimes |b\rangle_m&=&\delta_a^b\sum_{c=1}^4|c\rangle_l\otimes |c\rangle_m\,,
 \eea
 where $|a\rangle_l, a=1, \cdots, 4$ is the set of orthonormal basis of the Hilbert space at the $l$-th site. 
 	
	The $SU(4)$ alternating chain has four $R$ matrices,
	\begin{equation}
		\begin{aligned}& R_{0j}(\lambda)=\lambda\mathbb{I}+\mathbb{P}_{0j}\,, \\ & R_{0\bar{j}}(\lambda)=-(\lambda+2)\mathbb{I}+\mathbb{K}_{0\bar{j}}\,. \\ &R_{\bar{0}j}(\lambda)=-(\lambda+2)\mathbb{I}+\mathbb{K}_{\bar{0}j}\\
			& R_{\bar{0}\bar{j}}(\lambda)=\lambda\mathbb{I}+\mathbb{P}_{\bar{0}\bar{j}}.
		\end{aligned}
	\end{equation}
	Here, $0$ and $\bar{0}$ denote the auxiliary spaces which are the fundamental and anti-fundamental representation spaces of $SU(4)$ respectively. We define two monodromy matrices as
	\begin{align}
		&T_{0}(\lambda)=R_{01}(\lambda)R_{0\bar{1}}(\lambda)\cdots R_{0L}(\lambda)R_{0\bar{L}}(\lambda)\\ &\bar{T}_{\bar{0}}(\lambda)=R_{\bar{0}1}(\lambda)R_{\bar{0}\bar{1}}(\lambda)\cdots R_{\bar{0}L}(\lambda)R_{\bar{0}\bar{L}}(\lambda).
	\end{align}
	The transfer matrices are
	\begin{equation}\label{transfer}
		\tau(\lambda)=\mathrm{tr}_0T_0(\lambda),\quad\bar{\tau}(\lambda)=\mathrm{tr}_{\bar{0}}\bar{T}_{\bar{0}}(\lambda).
	\end{equation}
	A general Bethe state is characterized by three sets of Bethe roots, $\mathbf{u}$, $\mathbf{w}$, and $\mathbf{v}$, which satisfy the Bethe equations
	\begin{equation}\label{Bethe equations}
		\begin{gathered}
			1=\left(\frac{u_{j}+\frac{i}{2}}{u_{j}-\frac{i}{2}}\right)^{L}\prod_{k=1\atop k\neq j}^{N_{\mathbf{u}}}S(u_{j},u_{k})\prod_{k=1}^{N_{\mathbf{w}}}\tilde{S}(u_{j},w_{k}) \\
			1 =\prod\limits_{k=1\atop k\neq j}^{N_{\mathbf{w}}}S(w_{j},w_{k})\prod\limits_{k=1}^{N_{\mathrm{u}}}\tilde{S}(w_{j},u_{k})\prod\limits_{k=1}^{N_{\mathbf{v}}}\tilde{S}(w_{j},v_{k}) \\
			1 =\left(\frac{v_j+\frac{i}{2}}{v_j-\frac{i}{2}}\right)^L\prod\limits_{k=1\atop k\neq j}^{N_\mathbf{v}}S(v_j,v_k)\prod\limits_{k=1}^{N_\mathbf{w}}\tilde{S}(v_j,w_k), 
		\end{gathered}
	\end{equation}
	where $N_{\mathbf{u}}, N_{\mathbf{w}}, N_{\mathbf{v}}$ denote the number of rapidities in the sets $\mathbf{u}=\{u_1,\cdots,u_{N_{\mathbf{u}}}\},\mathbf{w}=\left\{w_1,\cdots,w_{N_{\mathbf{w}}}\right\}, \mathbf{v}=\{v_{1},\cdots,v_{N_{\mathbf{v}}}\}$,
	and
	\begin{equation}
		S(u,v)=\frac{u-v-i}{u-v+i},\quad\tilde{S}(u,v)=\frac{u-v+\frac{i}{2}}{u-v-\frac{i}{2}}.
	\end{equation}
	The cyclic property of single-trace operators also requires that the rapidities satisfy the zero-momentum condition
	\begin{equation}\label{zero momentum}
		1=\prod\limits_{j=1}^{N_{\mathbf{u}}}\frac{u_j+i/2}{u_j-i/2}\prod\limits_{j=1}^{N_{\mathbf{v}}}\frac{v_j+i/2}{v_j-i/2}.
	\end{equation}

 Another notation we need in the main part of the paper is the twisted integrable condition for the boundary state $|\mathcal{B}\rangle$
  	\be \tau(\lambda)|\mathcal{B} \rangle=
	\tau(-\lambda-2)|\mathcal{B} \rangle,\ee
	which leads to the following selection rules for the overlap $\langle \mathcal{B}|\mathbf{u}, \mathbf{w}, \mathbf{v}\rangle$ to be nonzero, 
	\be \mathbf{u}=-\mathbf{v}\,, \, \, \mathbf{w}=-\mathbf{w}\,.\ee

\section{Boundary state for odd $l_{pq|r}$ and special extremal $l_{pq|r}$}\label{appendix1}
When $l_{31|2}=0$, the boundary state is   
\begin{equation}
	\langle\mathcal{B}^{\mathrm{even}}_0 |=L \langle n_1@\emptyset, \bar{n}_1@\emptyset |
	\equiv L_3(\bar{n}_2)^{I_1}(n_2)_{J_1}\cdots (\bar{n}_2)^{I_{L_3}}(n_2)_{J_{L_3}}\langle I_1, J_1, \cdots, I_{L_3}, J_{L_3} |\,.
\end{equation}
The boundary state for the case $l_{31|2}=2L_3$ is 
\be\langle\mathcal{B}^{\mathrm{even}}_{2L_3} |=L \langle n_1@\{1, 2, \cdots, L_3\}, \bar{n}_1@\{ 1, 2, \cdots, L_3\} |\,.\ee
Then the result~\eqref{BoundaryEven} remains valid for the above two cases.

Now we turn to the case when $l_{31|2}$ is odd.  We similarly define,
\begin{eqnarray}
	&  \langle\bar{n}_1@\{1, 2,\cdots, m+1\},  n_1@\{1, 2, \cdots, m\}|=(\bar{n}_1)^{I_1}(n_1)_{J_1}\nonumber \\
	& \cdots (\bar{n}_1)^{I_m}(n_1)_{J_m}(\bar{n}_1)^{I_{m+1}}(n_2)_{J_{m+1}}\cdots (\bar{n}_2)^{I_L}(n_2)_{J_L}\langle I_1, J_1, \cdots, I_L, J_L|\,, \\
	& \langle\bar{n}_1@\{1, 2, \cdots, m\}, n_1@\{L, 1, 2, \cdots, m\} |=(\bar{n}_1)^{I_1}(n_1)_{J_1}\nonumber \\
	& \cdots (\bar{n}_1)^{I_m}(n_1)_{J_m}(\bar{n}_2)^{I_{m+1}}(n_2)_{J_{m+1}}\cdots (\bar{n}_2)^{I_L}(n_1)_{J_L}\langle I_1, J_1, \cdots, I_L, J_L |\, \\
	& \langle \mathcal{B}^{\mathrm{odd},\, a}_l|=\sum_{j=0}^{L-1}\langle \bar{n}_1@\{1, 2, \cdots, \frac{l+1}{2}\}, n_1@\{1, 2, \cdots, \frac{l-1}{2}\} |(U_{\mathrm{even}}U_{\mathrm{odd}})^j\,, \\
	& \langle\mathcal{B}_{l}^{\mathrm{odd}, \,b}|=\sum_{j=0}^{L-1}\langle \bar{n}_1@\{1, 2, \cdots, \frac{l-1}{2}\}, n_1@\{L, 1, 2, \cdots, \frac{l-1}{2}\} |(U_{\mathrm{even}}U_{\mathrm{odd}})^j\, , \\
	& \langle\mathcal{B}_{l}^{\mathrm{odd}}|=\langle\mathcal{B}_{l}^{\mathrm{odd}, \,a}|-\langle\mathcal{B}_{l}^{\mathrm{odd}, \,b}|\,.
\end{eqnarray}
From \eqref{WickOdd}, we have
\begin{align}
	\label{BoundaryOdd}
	\langle \hat{\mathcal{O}}_1^\circ(a_1) \hat{\mathcal{O}}_2^\circ(a_2) \hat{\mathcal{O}}_3(a_3)\rangle= \frac{(-1)^{\frac{l_{12|3}-1}{2}}\mathrm{sgn}(a_{12})}{N|a_1|^{l_{31|2}} |a_2|^{l_{23|1}}} L_1L_2\lambda^{\sum_{i=1}^3L_i}\langle\mathcal{B}^{\mathrm{odd}}_{l_{31|2}}  |\mathbf{u}, \mathbf{w}, \mathbf{v}\rangle\,. 
\end{align}
%\bea &&  \langle\bar{n}_1@\{ 2, 3, \cdots, l+1\}, , \bar{n}_1@\{1, 2, \cdots, l\}|=(\bar{n}_2)^{I_1}(n_1)_{J_1}(\bar{n}_1)_{I_2}(n_2)_{J_2}\nonumber \\
%&&\cdots (\bar{n}_1)^{I_l}(n_1)_{J_l}(\bar{n}_1)^{I_{l+1}}(n_1)_{J_{l+1}}\cdots (\bar{n}_2)^{I_1}(n_2)_{J_L}\langle I_1, J_1, \cdots, I_L, J_L| \eea
\section{Proof of $\mathcal{O}_{m, n}|\mathcal{B}_1^{\mathrm{odd,}\,b}\rangle=\mathcal{O}_{n, m}|\mathcal{B}_1^{\mathrm{odd,}\,b}\rangle$ for general $m, n$}\label{appendixA}

Let us prove that 
\be 
\mathcal{O}_{m, n}|\mathcal{B}_1^{\mathrm{odd,}\,b}\rangle=\mathcal{O}_{n, m}|\mathcal{B}_1^{\mathrm{odd,}\,b}\rangle\,,\label{eqappendix}\ee
for the most general case with $m,n\neq0$.
Using cyclicity of the  trace, each term in $\mathcal{O}_{m, n}$ can be written as %\footnote{{\color{red} The indices are a bit complicated. Examples from Peihe's note are complementary to the general discussions here. }}
\begin{eqnarray}\label{operator}
	&&\mathcal{O}^{\mathbb{P} \mathbb{K}}_{i^1_1\cdots {i^1_{a_1}}  {j^1_1}\cdots  {j^1_{b_1}}\cdots { i^s_1}\cdots {i^s_{a_s}} {j^s_1}, \cdots {j^s_{b_s}} }\\
	&=& {\mathrm tr}_0(\mathbb{P}_{0, {i^1_1}}\cdots\mathbb{P}_{0, {i^1_{a_1}}}\mathbb{K}_{0, {j^1_1}}\cdots\mathbb{K}_{0, { j^1_{b_1}}}\cdots \mathbb{P}_{0, {i^s_1}}\cdots\mathbb{P}_{0, {i^s_{a_s}}}\mathbb{K}_{0, {j^s_1}}\cdots\mathbb{K}_{0, {j^s_{b_s}}})\,,
\end{eqnarray}
with $1\leq s \leq L, 1\leq a_l \leq m, 1\leq b_l\leq n, \sum_{l=1}^{s}a_l=m$, $\sum_{l=1}^{s}b_l=n$.

The action of the above operator on the state
\begin{equation}|{I_{i^1_1}, \cdots, I_{i^1_{a_1}}}, {\bar{J}_{j^1_1}, \cdots, \bar{J}_{j^1_{b_1}}},\cdots,
	{I_{i^s_1}, \cdots, I_{i^s_{a_s}}}, {\bar{J}_{j^s_1}, \cdots, \bar{J}_{j^s_{b_s}}}\rangle\,,
\end{equation}
produces
\begin{eqnarray}
	% \nonumber % Remove numbering (before each equation)
	&& \delta_{{I_{i^1_1}}, \bar{J}_{j^s_{b_s}}}\delta_{I_{i^2_1}, {\bar{J}_{j^1_{b_1}}}}\cdots\delta_{{I_{i^s_1}}, \bar{J}_{j^{s-1}_{b_{s-1}}}}
	\delta_{A_{i^1_{a_1}}, \bar{B}_{j^1_1}} \cdots \delta_{A_{i^s_{a_s}}, \bar{B}_{j^s_1}}
	| {I_{i^1_2}, \cdots I_{i^1_{a_1}}}, A_{i^1_{a_1}}, \bar{B}_{j^1_1},  \\
	&&  {\bar{J}_{j_1^1}, \cdots, \bar{J}_{j^1_{b_1-1}}}, \cdots, { I_{i^s_2}, \cdots, I_{i^s_{a_s}}} A_{i^s_{a_s}}, \bar{B}_{j^s_1}, {\bar{J}_{j^s_1}, \cdots, \bar{J}_{j^s_{b_s-1}}} \rangle
\end{eqnarray}

Let us denote the state $\delta_{A_{i^1_{a_1}}, \bar{B}_{j^1_1}} |A_{i^1_{a_1}}, \bar{B}_{j^1_1}\rangle$ simply as
$|[i^1_{a_1},\bar{j}_1^1]\rangle$ with the underlying summation understood.

As a warm-up, we compute the action of $\mathcal{O}^{\mathbb{P} \mathbb{K}}_{\cdots}$ on $|\mathcal{B}^{\mathrm{even}}_0\rangle$,
\begin{eqnarray}
	& &\mathcal{O}^{\mathbb{P} \mathbb{K}}_{{i^1_1}\cdots {i^1_{a_1}}  {j^1_1}\cdots  {j^1_{b_1}}\cdots { i^s_1}\cdots {i^s_{a_s}} {j^s_1}, \cdots {j^s_{b_s}} }|\mathcal{B}_0^{\mathrm{even}}\rangle\nonumber\\&=&
	(n_2\cdot \bar{n}_2)^s|n_2\bar{n}_2 \cdots [i^1_{a_1} \small{|}\cdots\small{|} \bar{j}_1^1] n_2 \bar{n}_2 \cdots [i^s_{a_s} \small{|}\cdots\,\,\small{|} \bar{j}^s_1]\cdots
	\rangle\nonumber\\
	&:=& (n_2\cdot \bar{n}_2)^s| [i^1_{a_1} \bar{j}_1^1] \cdots [i^s_{a_s} \bar{j}^s_1]
	\rangle
\end{eqnarray}

Then we came back to the computations of $\mathcal{O}^{\mathbb{P} \mathbb{K}}_{\cdots}|\mathcal{B}^{{\mathrm{odd}}, \, b}_1\rangle$.
When $i\neq i^1_1, \cdots, i^1_{a_1}, \cdots, i^s_1, \cdots, i^s_{a_s}$, we have
\begin{eqnarray}
	&&  \mathcal{O}^{\mathbb{P} \mathbb{K}}_{{i^1_1}\cdots {i^1_{a_1}}  {j^1_1}\cdots  {j^1_{b_1}}\cdots {i^s_1}\cdots {i^s_{a_s}} {j^s_1}, \cdots {j^s_{b_s}} }|n_1@i\rangle\nonumber\\
	&=& (n_2\cdot \bar{n}_2)^s| n_1@i; [i^1_{a_1} \bar{j}_1^1] \cdots [i^s_{a_s} \bar{j}^s_1]
	\rangle
\end{eqnarray}

When $i=i^t_1$ for some $1\leq t\leq s$,
\begin{eqnarray}
	&&  \mathcal{O}^{\mathbb{P} \mathbb{K}}_{{i^1_1}\cdots {i^1_{a_1}}  {j^1_1}\cdots  {j^1_{b_1}}\cdots {i^s_1}\cdots {i^s_{a_s}} {j^s_1}, \cdots {j^s_{b_s}} }|n_1@i^t_1\rangle\nonumber\\
	&=& (n_2\cdot \bar{n}_2)^{s-1} (n_1 \cdot \bar{n}_2)|  [i^1_{a_1} \bar{j}_1^1] \cdots [i^s_{a_s} \bar{j}^s_1]
	\rangle
\end{eqnarray}
Notice that the $n_1$ in the ket operator has become $n_2$.
When $i=i^t_{m}, \,t=1, \cdots, s,\, m=2, 3, \cdots a_t$,
\begin{eqnarray}
	&&  \mathcal{O}^{\mathbb{P} \mathbb{K}}_{{i^1_1}\cdots {i^1_{a_1}}  {j^1_1}\cdots  {j^1_{b_1}}\cdots {i^s_1}\cdots {i^s_{a_s}} {j^s_1}, \cdots {j^s_{b_s}} }|n_1@i^t_m\rangle
 \nonumber\\
	&=& (n_2\cdot \bar{n}_2)^s| n_1@i^t_{m-1}; [i^1_{a_1} \bar{j}_1^1] \cdots [i^s_{a_s} \bar{j}^s_1]
	\rangle
\end{eqnarray}
Finally we get
\begin{eqnarray}\label{generalCase}
	&&  \mathcal{O}^{\mathbb{P} \mathbb{K}}_{{i^1_1}\cdots {i^1_{a_1}}  {j^1_1}\cdots  {j^1_{b_1}}\cdots { i^s_1}\cdots {i^s_{a_s}} {j^s_1}, \cdots {j^s_{b_s}} }|\mathcal{B}^{{\mathrm{odd}}, \, b}_1\rangle\nonumber\\
	&=& s (n_2\cdot \bar{n}_2)^{s-1} (n_1 \cdot \bar{n}_2)|  [i^1_{a_1} \bar{j}_1^1] \cdots [i^s_{a_s} \bar{j}^s_1]\rangle\nonumber\\
	&+&\sum_{i\neq i^t_{a_t}}(n_2\cdot \bar{n}_2)^s| n_1@i; [i^1_{a_1} \bar{j}_1^1] \cdots [i^s_{a_s} \bar{j}^s_1]\rangle\,.
\end{eqnarray}
Notice that the result only depends on $s$ and  $i^1_{a_1}, \bar{j}_1^1, \cdots, i^s_{a_s}, \bar{j}^s_1$.

We group the terms in $\mathcal{O}_{m, n} $ that share the same parameters $s, i^1_{a_1}, j_1^1, \cdots, i^s_{a_s}, j^s_1$. Denote this collection of terms as 
$X_{s, \{i^t_{a_t},j^t_1\}}$.

 In $\mathcal{O}_{n, m}$, we consider the terms  satisfied the constraints $s^\prime=s, a_i^\prime=b_{i-1}, b_i^\prime=a_{i+1}, i^{\prime t}_{a^\prime_t}=i^t_{a_t}$ and $j^{\prime t}_1=j^t_1$. Call the collection of these terms $X^\prime_{s, \{i^{t}_{a_t}, j^t_1\}}$.
Notice that for these operators, the constraints $\sum_{l=1}^s a_l^\prime=\sum_{l=1}^sb_{l-1}=n,  \sum_{l=1}^s b_l^\prime=\sum_{l=1}^sa_{l+1}=m$
are satisfied.

From~\eqref{generalCase}, we know that each term in  the collection $X_{s, \{i^t_{a_t}, j^t_1\}}$ or $X^\prime_{s, \{i^{t}_{a_t}, j^t_1\}}$
yields the same results when acting on $|\mathcal{B}^{{\mathrm odd}, \, b}_1\rangle$. To prove \eqref{criterion}, we need to demonstrate that these two collections contain exactly the same number of terms.

Let us first count the number of ways of putting $\mathbb{K}_{0, j_2^1}, \cdots, \mathbb{K}_{0, j_{b_1^1}},  \mathbb{P}_{0, i^2_1}, \cdots, \mathbb{P}_{0, i^2_{a_2-1}} $
between $\mathbb{K}_{0, j_1^1}$ and $\mathbb{P}_{0, i^2_{a_2}} $. We need to choose $a_2-1+b_1-1=b_1+a_2-2$ positions for these $\mathbb{K}$'s and $\mathbb{P}$'s  among $i^2_{a_2}-j^1_1-1$ positions between $\mathbb{K}_{0, j_1^1}$ and $\mathbb{P}_{0, i^2_{a_2}}$.
Since  that all $\mathbb{K}$'s should be to the left of all $\mathbb{P}$'s, their relative order does not introduce additional factors. Therefore, the total number of choices is given by
$C_{i^2_{a_2}-j^1_1-1}^{b_1+a_2-2}$.

Because the placement choices between
$\mathbb{K}_{0, j_1^t}$ and $\mathbb{P}_{0, i^{t+1}_{a_{t+1}}}$ are independent for different $t$,
the  number of terms in $X_{s, \{i^t_{a_t}, j^t_1\}}$ equals to
\begin{equation}
	N_X= \prod_{t=1}^s \binom{i^t_{a_t}-j^{t-1}_1-1}{b_{t-1}+a_t-2}\,.
\end{equation}

This counting also leads to the result that the number of terms in $X^\prime_{s, \{i^{t}_{a_t}, j^t_1\}}$ is
\be
% \nonumber % Remove numbering (before each equation)
N_{X^\prime} =\prod_{t=1}^{s^\prime}\binom{i^{\prime t}_{a^\prime_t}-j^{\prime t-1}_1-1}{b^\prime_{t-1}+a^\prime_t-2}  \,.
\ee
Using the constraints
\be s^\prime=s\,,\, a_i^\prime=b_{i-1}\,,\, b_i^\prime=a_{i+1}\,,\,i^{\prime t}_{a^\prime_t}=i^t_{a_t}\,,\,j^{\prime t}_1=j^t_1\,,  \ee
we get 
\be  N_{X^\prime}=N_X\,.\ee
This finishes our proof of \eqref{eqappendix}
for general $m, n$.


\begin{thebibliography}{10}

\bibitem{Aharony:2008ug}
Ofer Aharony, Oren Bergman, Daniel~Louis Jafferis, and Juan Maldacena.
\newblock {${\mathcal N}=6$ superconformal Chern-Simons-matter theories,
  M2-branes and their gravity duals}.
\newblock {\em JHEP}, 10:091, 2008.

\bibitem{Bak:2008cp}
Dongsu Bak and Soo-Jong Rey.
\newblock {Integrable Spin Chain in Superconformal Chern-Simons Theory}.
\newblock {\em JHEP}, 10:053, 2008.

\bibitem{Basso:2022nny}
Benjamin Basso, Alessandro Georgoudis, and Arthur~Klemenchuk Sueiro.
\newblock {Structure Constants of Short Operators in Planar ${\mathcal N}=4$
  Supersymmetric Yang-Mills Theory}.
\newblock {\em Phys. Rev. Lett.}, 130(13):131603, 2023.

\bibitem{Basso:2015zoa}
Benjamin Basso, Shota Komatsu, and Pedro Vieira.
\newblock {Structure Constants and Integrable Bootstrap in Planar ${\mathcal
  N}=4$ SYM Theory}.
\newblock 5 2015.

\bibitem{Beisert:2010jr}
Niklas Beisert et~al.
\newblock {Review of AdS/CFT Integrability: An Overview}.
\newblock {\em Lett. Math. Phys.}, 99:3--32, 2012.

\bibitem{Bissi:2012ff}
Agnese Bissi, Charlotte Kristjansen, Ara Martirosyan, and Marta Orselli.
\newblock {On Three-point Functions in the $AdS_4/CFT_3$ Correspondence}.
\newblock {\em JHEP}, 01:137, 2013.

\bibitem{Bombardelli:2017vhk}
Diego Bombardelli, Andrea Cavagli\`a, Davide Fioravanti, Nikolay Gromov, and
  Roberto Tateo.
\newblock {The full Quantum Spectral Curve for $AdS_4/CFT_3$}.
\newblock {\em JHEP}, 09:140, 2017.

\bibitem{Cavaglia:2014exa}
Andrea Cavagli\`a, Davide Fioravanti, Nikolay Gromov, and Roberto Tateo.
\newblock {Quantum Spectral Curve of the ${\mathcal N}=6$ Supersymmetric
  Chern-Simons Theory}.
\newblock {\em Phys. Rev. Lett.}, 113(2):021601, 2014.

\bibitem{deLeeuw:2015hxa}
Marius de~Leeuw, Charlotte Kristjansen, and Konstantin Zarembo.
\newblock {One-point Functions in Defect CFT and Integrability}.
\newblock {\em JHEP}, 08:098, 2015.

\bibitem{DHoker:1999jke}
Eric D'Hoker, Daniel~Z. Freedman, Samir~D. Mathur, Alec Matusis, and Leonardo
  Rastelli.
\newblock {Extremal correlators in the AdS / CFT correspondence}.
\newblock pages 332--360, 8 1999.

\bibitem{Drukker:2009sf}
Nadav Drukker and Jan Plefka.
\newblock {Superprotected n-point correlation functions of local operators in
  ${\mathcal N}=4$ super Yang-Mills}.
\newblock {\em JHEP}, 04:052, 2009.

\bibitem{Escobedo:2010xs}
Jorge Escobedo, Nikolay Gromov, Amit Sever, and Pedro Vieira.
\newblock {Tailoring Three-Point Functions and Integrability}.
\newblock {\em JHEP}, 09:028, 2011.

\bibitem{Gombor:2020kgu}
Tamas Gombor and Zoltan Bajnok.
\newblock {Boundary states, overlaps, nesting and bootstrapping AdS/dCFT}.
\newblock {\em JHEP}, 10:123, 2020.

\bibitem{Gromov:2013pga}
Nikolay Gromov, Vladimir Kazakov, Sebastien Leurent, and Dmytro Volin.
\newblock {Quantum Spectral Curve for Planar $\mathcal{N} = 4$ Super-Yang-Mills
  Theory}.
\newblock {\em Phys. Rev. Lett.}, 112(1):011602, 2014.

\bibitem{Gromov:2014caa}
Nikolay Gromov, Vladimir Kazakov, S\'ebastien Leurent, and Dmytro Volin.
\newblock {Quantum spectral curve for arbitrary state/operator in
  $AdS_{5}/CFT_{4}$}.
\newblock {\em JHEP}, 09:187, 2015.

\bibitem{Hutsalyuk:2024saw}
Arthur Hutsalyuk, Yunfeng Jiang, Balazs Pozsgay, Hefeng Xu, and Yang Zhang.
\newblock {Exact Spin Correlators of Integrable Quantum Circuits from Algebraic
  Geometry}.
\newblock 5 2024.

\bibitem{Ivanovskiy:2024vel}
Vyacheslav Ivanovskiy, Shota Komatsu, Victor Mishnyakov, Nikolay Terziev,
  Nikita Zaigraev, and Konstantin Zarembo.
\newblock {Vacuum Condensates on the Coulomb Branch}.
\newblock 5 2024.

\bibitem{JKV}
Yunfeng Jiang, Shota Komatsu, and Edoardo Vescovi.
\newblock {to appear}.

\bibitem{Jiang:2019zig}
Yunfeng Jiang, Shota Komatsu, and Edoardo Vescovi.
\newblock {Exact Three-Point Functions of Determinant Operators in Planar
  ${\mathcal N}=4$ Supersymmetric Yang-Mills Theory}.
\newblock {\em Phys. Rev. Lett.}, 123(19):191601, 2019.

\bibitem{Jiang:2019xdz}
Yunfeng Jiang, Shota Komatsu, and Edoardo Vescovi.
\newblock {Structure constants in $ \mathcal{N} = 4$ SYM at finite coupling as
  worldsheet g-function}.
\newblock {\em JHEP}, 07(07):037, 2020.

\bibitem{Jiang:2023cdm}
Yunfeng Jiang, Jun-Bao Wu, and Peihe Yang.
\newblock {Wilson-loop one-point functions in ABJM theory}.
\newblock {\em JHEP}, 09:047, 2023.

\bibitem{Kazama:2014sxa}
Yoichi Kazama, Shota Komatsu, and Takuya Nishimura.
\newblock {Novel construction and the monodromy relation for three-point
  functions at weak coupling}.
\newblock {\em JHEP}, 01:095, 2015.
\newblock [Erratum: JHEP 08, 145 (2015)].

\bibitem{Kristjansen:2021abc}
Charlotte Kristjansen, Dinh-Long Vu, and Konstantin Zarembo.
\newblock {Integrable domain walls in ABJM theory}.
\newblock {\em JHEP}, 02:070, 2022.

\bibitem{Kristjansen:2023ysz}
Charlotte Kristjansen and Konstantin Zarembo.
\newblock {\textquoteright{}t Hooft loops and integrability}.
\newblock {\em JHEP}, 08:184, 2023.

\bibitem{Minahan:2008hf}
J.~A. Minahan and K.~Zarembo.
\newblock {The Bethe ansatz for superconformal Chern-Simons}.
\newblock {\em JHEP}, 09:040, 2008.

\bibitem{Pereira:2017unx}
Raul Pereira.
\newblock {\em {Correlation Functions in Integrable Theories: From weak to
  strong coupling}}.
\newblock PhD thesis, Uppsala U., 2017.

\bibitem{Piroli:2017sei}
Lorenzo Piroli, Bal\'azs Pozsgay, and Eric Vernier.
\newblock {What is an integrable quench?}
\newblock {\em Nucl. Phys. B}, 925:362--402, 2017.

\bibitem{Simmons-Duffin:2016gjk}
David Simmons-Duffin.
\newblock {The Conformal Bootstrap}.
\newblock In {\em {Theoretical Advanced Study Institute in Elementary Particle
  Physics}: {New Frontiers in Fields and Strings}}, pages 1--74, 2017.

\bibitem{Sutherland:1975vr}
Bill Sutherland.
\newblock {A General Model for Multicomponent Quantum Systems}.
\newblock {\em Phys. Rev. B}, 12:3795--3805, 1975.

\bibitem{1970JETPL..12..225U}
G.~V. {Uimin}.
\newblock {One-dimensional Problem for S = 1 with Modified Antiferromagnetic
  Hamiltonian}.
\newblock {\em JETP Lett.}, 12:225, January 1970.

\bibitem{Yang:2022dlk}
Peihe Yang.
\newblock {Integrable boundary states from maximal giant gravitons in ABJM
  theory}.
\newblock {\em Phys. Lett. B}, 846:138194, 2023.

\bibitem{Yang:2021hrl}
Peihe Yang, Yunfeng Jiang, Shota Komatsu, and Jun-Bao Wu.
\newblock {Three-point functions in ABJM and Bethe Ansatz}.
\newblock {\em JHEP}, 01:002, 2022.

\end{thebibliography}
\end{document}